\documentstyle[12pt]{article}
\newlength{\extraspace}
\newlength{\extraspaces}
\newlength{\zpaces}
\newlength{\zpace}
\renewcommand{\l}{$l$}

\newcommand{\nd}{$\nu$}

\newcommand{\n}{\nu}

\newcommand{\rmn}{r^{-|\n|}}

\newcommand{\rpln}{r^{|l+\n|}}

\newcommand{\rplmn}{r^{|l+\n|-|\n|}}

\renewcommand{\u}{\Psi}

\renewcommand{\v}{\tilde{\Psi}}

\newcommand{\hv}{\tilde{H}}

\newcommand{\be}{\begin{equation}
\addtolength{\abovedisplayskip}{\extraspaces}
\addtolength{\belowdisplayskip}{\zpaces}
\addtolength{\abovedisplayshortskip}{\extraspace}
\addtolength{\belowdisplayshortskip}{\zpace}}
\newcommand{\ee}{\end{equation}}
\newcommand{\bea}{\begin{eqnarray}
\addtolength{\abovedisplayskip}{\extraspaces}
\addtolength{\belowdisplayskip}{\zpaces}
\addtolength{\abovedisplayshortskip}{\extraspace}
\addtolength{\belowdisplayshortskip}{\zpace}}
\newcommand{\eea}{\end{eqnarray}}
\newcommand{\newsection}[1]{
\pagebreak[3]
\setcounter{equation}{0}
\setcounter{footnote}{0}
\section{#1}
\nopagebreak
\medskip
\nopagebreak}
\newcommand{\newsubsection}[1]{
\subsection{#1}}
\newcommand{\ber}{\begin{eqnarray}}
\newcommand{\eer}{\end{eqnarray}}

\newcounter{tabnum}
\font\twelvebf=cmbx10 scaled\magstep 1
\begin{document}
\newlength{\gacspazio}
\setlength{\gacspazio}{5.2cm}
\newlength{\espacio}
\setlength{\espacio}{3.1mm}

\newpage
\centerline{{\bf PERTURBATIVE RENORMALIZATIONS OF  }}
\centerline{{\bf ANYON QUANTUM MECHANICS}\footnote{This work is
supported in part by funds provided by the U.S. Department of Energy (D.O.E.)
under cooperative agreement \#DE-FC02-94ER40818,
and by the Istituto Nazionale di
Fisica Nucleare (INFN, Frascati, Italy).}}
\vskip 50pt
\bigskip
\centerline{Giovanni AMELINO-CAMELIA}
\vskip 18pt
\baselineskip 12pt plus 0.2pt minus 0.2pt
\centerline{{\it Center for Theoretical Physics}}
\centerline{{\it Laboratory for Nuclear Science}}
\centerline{{\it and Department of Physics}}
\centerline{{\it Massachusetts Institute of Technology}}
\centerline{{\it Cambridge, Massachusetts 02139 USA}}
%
%
\vskip 3cm
\centerline{\bf ABSTRACT }
In bosonic end perturbative calculations for quantum mechanical
anyon  systems a regularization and renormalization procedure,
analogous to those used in field theory, is necessary.
I examine the reliability and the physical interpretation of
the most commonly used
bosonic end regularization procedures.
I then use the regularization procedure with the most transparent
physical interpretation to derive some bosonic end perturbation
theory results on anyon spectra, including a 3-anyon ground state
energy.


\vskip 2cm
\centerline{Submitted to: Physical Review D}
\vfill
\noindent{MIT-CTP-2321 \hfill June, 1994}

\newpage
\baselineskip 12pt plus 0.2pt minus 0.2pt
\newsection{Introduction}
In 2+1 dimensions the rotation group is abelian ($SO(2)$), and
as a consequence,
besides bosons and fermions,
also particles with neither integer nor half-integer spin, anyons,
can exist\cite{wil}.
Free anyons can be described\cite{jackpi}, in what is called
``boson gauge" (or ``magnetic gauge") description, as
non-relativistic bosons\footnote{Indeed, in 2+1 dimensions particles with
any statistics can be described as bosons interacting through an appropriate
statistical interaction.}
interacting with each other
through the mediation of an
abelian Chern-Simons gauge field; the corresponding Hamiltonian is
\bea
H_b = {1 \over 2 m} \sum_n \biggl( {\bf p}_n - \nu {\bf a}_n \biggr)^2
{}~, \label{hany}\\
a_n^k \equiv - \epsilon^{kj} \sum_{m (\ne n)} {r_n^j - r_m^j \over
|{\bf r}_n - {\bf r}_m|^2 }
{}~, \label{amatt}
\eea

\noindent
where ${\bf r_n} \equiv (r_n^1, r_n^2)$
is the position vector of the n-th particle.

An alternative description of anyons, called ``anyon gauge" description,
can be derived from the boson gauge using the following transformation
\be
\Psi_b ~ \rightarrow ~ \Psi_a ~ = ~ U ~ \Psi_b ~,~~~
H_b ~ \rightarrow ~ H_a~ = ~ U ~ H_b ~ U^{-1}
{}~, \label{tratra}
\ee

\noindent
where
\be
U \equiv \exp \biggl[ - i \nu \sum_{m \ne n} \theta_{n m} \biggr] ~,
\label{utra}
\ee

\noindent
the $\Psi_b$'s are the bosonic wave functions that appear in the boson
gauge description, and $\theta_{n m}$ are the azimuthal angles of the
relative vectors ${\bf r}_n - {\bf r}_m$.
It is easy to verify that $H_a$ has the
form $\sum_{n} {\bf p}_n^2/2m$ of an ordinary free Hamiltonian;
moreover, since the boson gauge wave functions
$\Psi_b$ are single-valued and $U$ is not single-valued,
the $\Psi_a$'s are multi-valued.
Indeed, with the transformation (\ref{tratra}) one removes the interaction
by redefining (in a multivalued fashion) the phase of the wave functions,
a procedure which is made possible by the fact that ${\bf a_n}$ is
a (singular) pure gauge potential\cite{jackpi}.
The multi-valuedness of the $\Psi_a$'s is
directly related to the anomalous quantum
statistics of anyons, which, in particular, prescribes\cite{wil}
that the quantum mechanical wave functions
describing the relative motion
of two anyons in polar coordinates\footnote{Here the relative angle
runs from $- \infty$ to $\infty$, without identifying
angles which differ by multiples of $2 \pi$,
so that it keeps track of the number of windings\cite{wil}.},
satisfy the following condition
\be
\Psi_a(r,\theta+\pi) = e^{- i \nu \pi} \Psi_a(r,\theta)
{}~.
\label{anystat}
\ee

\noindent
It is easy to verify that the $\Psi_a$'s defined in
Eq.(\ref{tratra}) satisfy the condition (\ref{anystat}).

The parameter $\nu$, called ``statistical parameter",
characterizes the type of
anyons, i.e. their statistics.
In particular, from Eq.(\ref{anystat}) one realizes that anyons with
even (odd) integer $\nu$ verify bosonic (fermionic) statistics, whereas
the noninteger values of $\nu$ correspond to particles with
statistics interpolating between the bosonic and the fermionic case.
Without any loss of generality\cite{wil},
I shall restrict the values of $\nu$ to be in the interval
[-1,1].

Whereas in the case of
bosons and fermions the wave functions for $N$ non-interacting particles
or $N$ particles interacting via a separable Hamiltonian (i.e. an Hamiltonian
which can be written as a sum of corresponding one-particle Hamiltonians)
are simply given by appropriately symmetrized products of one-particle
wave functions, in the case of
general-type anyons, because of the anomalous statistics,
these factorizations do not occur, rendering the study of any
$N$-anyon problem extremely difficult.
It is essentially for this reason that the problem of finding
all the exact energy eigenvalues and
eigenfunctions, even for very simple
anyon systems, is still unsolved for $N>2$.

The realization that anyons are
relevant to the understanding of some condensed matter phenomena,
most notably the fractional quantum Hall effect\cite{lh},
has motivated numerous recent studies of
the $N$-anyon problem.
In particular, the $N$-identical-anyon in an external harmonic potential
system has
played a central role in these recent investigations.
The harmonic potential, by discretizing the spectrum,
allows to disentangle the dependence on the statistical parameter $\nu$;
moreover, the virial coefficients can be deduced from the solutions
of the harmonic potential problem by taking in appropriate
fashion\footnote{Note that the harmonic oscillator potential
has the same form as one of the generators of the $SO(2,1)$ symmetry
transformations of the system\cite{jakso21}.
It would be interesting to investigate whether this fact plays a role
in the various uses made of the harmonic oscillator potential
in the study of anyon quantum mechanics, especially concerning
the limiting procedure to be followed in deriving the virial coefficients.}
the limit of vanishing
oscillator frequency\cite{verbapert,senprd}.

The $N$-anyon in harmonic oscillator problem has been completely
solved\cite{wil} for $N\!=\!2$, but only
an incomplete set of exact eigensolutions
has been found\cite{wu,tru,cho} for $N>2$.
Interestingly all the known
eigenenergies depend linearly  on the statistical
parameter $\nu$.

Besides the search for exact solutions,
there has been much effort in the
perturbati- ve[5,6,9-16]
and in the numerical\cite{verba,murthy} study of the
$N$-anyon problem.
Both the perturbative and the numerical studies have led to interesting
results; most notably, they have shown that there are energies with nonlinear
dependence on $\nu$ among those that are not presently known exactly.

In the proposed perturbative approaches,
called ``bosonic end" and ``fermionic end" perturbative
approaches, one studies either
``quasi-bosonic" anyons (i.e. anyons with small $\nu$,
whose anomalous statistics
is close to being bosonic) or ``quasi-fermionic" anyons,
by using a perturbative expansion
in which the small parameter is the deviation of the statistics
(indicated by the statistical parameter $\nu$) from the bosonic
or the fermionic limit.

The analysis presented in this paper concerns the
bosonic end perturbative approach,
which, because of the non-analyticity\cite{wil,gac}
of the limit $\nu \rightarrow 0$, is afflicted by some spurious
divergencies, and therefore requires
regularization[9,11,16,19-21].
In order to see the mechanism that leads to these divergencies,
let us look at the simple case
of two anyons in an harmonic oscillator
potential; the boson gauge Hamiltonian
that describes the relative motion is
\be
H_2 = - {1 \over r} \partial_{r} (r \partial_{r})
- {1 \over r^2}  \partial_{\phi}^{2}
+ r^2
-{ 2 i \nu \over  r^2}\partial_{\phi} + { \nu^2 \over r^2} ~.
\label{eqbe}
\ee

\noindent
In the perturbative calculations about $\nu=0$,
one runs into
an inconsistency because the matrix elements of
${\nu^2 \over r^2}$ between 0-th
order in $\nu$ (i.e. bosonic) s-wave functions
are logarithmically divergent; for example, to the second order in
$\nu$, one of the contributions to the ground state energy is given by
\bea
<\Omega_{boson}| {\n^2 \over r^2} |\Omega_{boson}> &=&
\int^{\infty}_{0} \int^{2\pi}_{0} r~dr~d\phi
{}~{e^{ - {r^2  \over 2}} \over \pi^{1 \over 2}}~
{\n^2 \over r^2}
{}~{e^{ - {r^2  \over 2}} \over \pi^{1 \over 2}}~     \nonumber\\
&=& 2 \nu^2 \int^{\infty}_{0} {\exp( -r^2 ) \over r}~dr~
\sim \infty ~.
\label{eqbf}
\eea

Many of the results presently available on
the unknown portion of the anyon spectra,
have been obtained using
bosonic end perturbation theory.
However,
some of the regularization
procedures used in these calculations
require rather arbitrary manipulations.
Only recently, motivated by the results of some
related field theory analysis\cite{lozanob}, a regularization procedure
with a clearer physical interpretation has been
proposed\cite{pap9,manu,ouvnew}.
This regularization procedure is based on the introduction
of a repulsive $\delta$-function potential, and I shall refer to it
as the ``$\delta$-function regularization".

The plan of this paper is the following. In the next section, I review
the three most commonly\footnote{A fourth
regularization procedure has been discussed in Ref.\cite{ouvnew};
however, that regularization procedure has not been used
in the study of anyon spectra, and in Ref.\cite{ouvnew}
it was shown to be completely equivalent to one of the
three regularization procedures discussed in the following.}
used procedures of regularization of bosonic
end perturbation theory. In particular, this allows me to show
that, among these procedures, the $\delta$-function regularization
is indeed the one with the clearest physical interpretation.
In Sec.III, I discuss the relations among the different regularization
procedures.
In Sec.IV, I explicitly test\footnote{Most of these tests
have already been performed in the short publication \cite{pap9};
in Sec.IV I review them, giving additional details on the calculations.}
the reliability of the $\delta$-function regularization by using it in
rederiving several results that have already been obtained exactly
or numerically.
Finally, Sec.V is devoted to my conclusions.

\newsection{Regularizations of Bosonic End Perturbation Theory}
In this
section I briefly review the three most commonly
used procedures
of regularization of bosonic end perturbation theory.
In order to be specific, I limit the discussion to the case of 2 identical
anyons in an
external harmonic oscillator potential, whose relative motion is described
by the Hamiltonian $H_2$ in Eq.(\ref{eqbe}).
As I mentioned, the exact solutions of
the $H_2$-eigenproblem are known;
they are given by\cite{wil}
\be
E^{exact}_{n,l,\nu} = ( 4n + 2|l+\n| + 2)
{}~, \label{solb}
\ee
\be
|\u_{n,l,\nu}^{exact}> =
N_{n,l}^{\nu}~ \rpln ~
e^{ - {r^2 \over 2} + i l \phi }~
L_{n}^{|l+\nu|}(r^2)
{}~. \label{sola}
\ee

\noindent
where the $L_{n}^{x}$ are
Laguerre polynomials, and the $N_{n,l}^{\nu}$ are normalization
constants.

\newsubsection{Regularization Method I}
It has been proposed\cite{comtet,sen} that, in order
to obtain a consistent
bosonic end perturbative approach,
one should not apply perturbation theory
to the original eigenproblem
$H_N |\Psi> = E |\Psi>$ (where $H_N$ is a $N$-anyon Hamiltonian),
but rather
to the modified eigenproblem $\hv_N |\v> =E |\v>$
where $\hv^\nu_N$ and $|\v>$
are related to the original $H_N$ and $|\Psi>$ by
\be
|\v> = \left(\prod_{n<m} ({\bf r_{n} - r_{m}})^{-|\nu|} \right) |\u>
\label{eqcaextra}
\ee
\be
\hv^{\nu}_N = \left(\prod_{n<m} ({\bf r_{n} - r_{m}})^{-|\nu|} \right) H_N
\left(\prod_{n<m} ({\bf r_{n} - r_{m}})^{|\nu|} \right) ~.
\label{eqca}
\ee

\noindent
In the case of two anyons
in an external harmonic oscillator potential,
from (\ref{eqbe}) one finds that the Hamiltonian is
\be
\hv^{\nu}_2 = r^{- |\nu|} H_2 r^{|\nu|}
={1 \over r} \partial_{r} (r \partial_{r})
- {1 \over r^2}  \partial_{\phi}^{2} + r^2
 -{ 2i\n \over  r^2}\partial_{\phi}
- {2 |\n| \over  r}\partial_{r}.
\label{eqcb}
\ee

\noindent
The exact solutions for the
$\hv_2^{\nu}$-eigenproblem are easily found from
(\ref{solb}) and (\ref{sola}):
\be
{\tilde E}^{exact}_{n,l,\nu} = E^{exact}_{n,l,\nu}
= ( 4n + 2|l+\n| + 2)
\label{eqcca}
\ee
\be
|\v_{n,l,\nu}^{exact}> = {{\tilde{N}}_{n,l}^{\n} \over {{N}}_{n,l}^{\n}}
r^{- |\nu|} |\Psi_{n,l,\nu}^{exact}> =
{\tilde{N}}_{n,l}^{\n}~\rplmn ~
\exp\left( - {r^2 \over 2} + i l \phi \right)~
L_{n}^{|l+\n|}(r^2)
\label{eqccb}
\ee

\noindent
where  ${\tilde{N}}_{n,l}^{\n}~$ are normalization constants.

Notice that: (i)\space\space $\hv_2^{\nu}$ is not Hermitian,
(ii)\space\space the transformation $|\u> \rightarrow |\v> \sim \rmn |\u>$
is not unitary,
(iii)\space\space $\hv_2^{\nu}$ depends non-analytically
on \nd.
Moreover, the eigensolutions (i.e. the eigenfunctions and the eigenvalues)
of $\hv_2^{\nu}$ are non-smooth functions
of \nd\space
for every value of \l,
whereas the $H_2$-eigensolutions are
non-smooth only for \l=0.

Although the exact $H_2$-eigenproblem and
$\hv_2^{\nu}$-eigenproblem are simply related,
the same is not true when perturbation theory is used.
In particular, the non-analytic dependence on $\nu$ present in
$\hv_2^{\nu}$ is such to compensate (in perturbation theory) for the
other non-analitycities present in the problem, and, as a consequence,
the bosonic end perturbation theory for the
$\hv_2^{\nu}$-eigenproblem is not affected by divergencies\cite{comtet,gac}.

The ``regularization method I" consists in using
the (finite) results of perturbation theory applied to
the $\hv^{\nu}_N$-eigenproblem to derive, based on
the relation between $|\v>$ and $|\u>$,
the corresponding results for the
original $H_N$-eigenproblem.

This regularization method I has undergone
numerous tests\cite{senprd,comtet,gac}.
For the 2-anyon in harmonic potential problem
all the regularization method I results
have been shown to be consistent with the
exact solutions.
Moreover, several first order
few-anyon eigenenergies have been calculated
and the results are in agreement with the
expansion in $\nu$ of the exact solutions.

Unfortunately,
the physics behind the manipulations involved in
the regularization method I
is not completely understood.
For example, it is not clear
for what reason among the
Hamiltonians $\hv_2^\beta \equiv r^{-|\beta|} ~ H_2 ~ r^{|\beta|}$
only the one with $\beta=\nu$ leads to the correct
anyon spectra.
It would also be interesting to comprehend the role in the regularization
procedure
of the fact that
the transformation $|\u> \rightarrow |\v>$ is
singular at $r=0$ and non-unitary,
and that $\hv^{\nu}_N$ is not Hermitian.
An ``hermitized version'' of the regularization method I,
in which the perturbation theory is based on the Hermitian Hamiltonian
$(\hv^{\nu +}_N + \hv^{\nu}_N)/2$, has also been considered in the
literature\cite{sen,kimkwon}, but it can be used only for rather
limited tasks\cite{sen,gac} (more on this in Sec.III).

\newsubsection{Regularization Method II}
Another perturbative approach to the study of quasi-bosonic anyons
used in the literature\cite{cho,mit,papsix} is based on the idea
that, if one wants to describe perturbatively
the conventional ``non-colliding" anyons\footnote{``Non-colliding anyons"
are anyons whose wave functions vanish at the points of overlap.
The possibility of ``colliding" anyons has
also been examined in the literature
(for example this subject is discussed in
Ref.\cite{bourdeau}), but in this paper only
the conventional ``non-colliding'' anyons are considered.}
in terms of (0-th order)
bosonic wave functions, one
needs to modify the bosonic
wave functions so that they have an appropriate hard core.
Specifically, one substitutes
the bosonic
wave functions $|\u^{(0)}>$ with ``regularized bosonic wave
functions" $|\u^{(0)R}>$
in all the divergent matrix elements
of the naive perturbative approach (like the matrix element
in Eq.(\ref{eqbf})).
For the 2-body case (the generalization
to the N-body case is given in the references)
the $|\u^{(0)R}>$ are defined by
\be
|\u^{(0)R {\nu \over 2}}_{n,l}>
\equiv {{N}}_{n,l}^{R ~ \n}~r^{{|\n| \over 2}}~|\u^{(0)}_{n,l}>
{}~, \label{eqcq}
\ee

\noindent
where ${{N}}_{n,l}^{R ~ \n}$ are normalization constants.
Eq.(\ref{eqcq}) indeed implements
a hard core condition; in fact $|\u^{(0)R {\nu \over 2}}_{n,l}> = 0$ for $r=0$.
As a consequence, substituting the $|\u^{(0)R {\nu \over 2}}_{n,l}>$ to the
$|\u^{(0)}_{n,l}>$ in the divergent matrix elements of the naive
perturbative approach, one obtains finite
matrix elements; for example the matrix element in (\ref{eqbf}) is
substituted by
\bea
<\u^{(0)R{\nu \over 2}}_{0,0}|{\n^2  \over  r^2}
|\u^{(0)R{\nu \over 2}}_{0,0}>
\!\!&=&\!\! \int^{\infty}_{0} \int^{2 \pi}_{0}
{r~dr~d\phi~ \over \pi \Gamma(1+{|\n| \over 2}) }~
{\n^2  \over  r^2} ~e^{-{r^2 }}~r^{|\n|} \nonumber\\
\!\!&=&\!\! \n^2  {\Gamma({|\n| \over 2})
\over \Gamma(1+{|\n| \over 2})}
= 2 |\n|
{}~. \label{eqcr}
\eea

\noindent
The choice of the ``regularizing exponent" ${|\n| \over 2}$ in (\ref{eqcq}),
is only justified {\it a posteriori}
by the agreement of the perturbative results with the exact results\cite{gac}.

It is important to notice that, although (\ref{eqcq}) and (\ref{eqccb})
are formally
similar, this ``regularization method II" and the
regularization method I
represent conceptually different approaches.
In the
regularization method I,
one studies a different eigenproblem, since
the Hamiltonian $\hv_N$
is different from $H_N$, and
perturbation theory is applied to the eigenproblem of $\hv_N$;
only at the end
one recovers a perturbative result for the original eigenproblem through
appropriate conversion formulas\cite{gac}.
In the regularization method II,
equation (\ref{eqcq}), which involves only the 0-th order
wave functions,
effectively defines regularized matrix elements for the original theory,
and no equivalent eigenproblem is introduced; in the
calculations one uses the original Hamiltonian,
and the results refer directly (no conversion formulas are needed) to the
original eigensolutions.

Another important observation is
that in Eq.(\ref{eqcr}) a matrix element apparently of order $\nu^2$
actually gives a contribution of order $\nu$.
This is a general aspect of the
regularization method I, and
is due to the fact that near $\nu=0$ some matrix elements
$<\Psi_0^{R{\nu \over 2}}| r^{-2} |\Psi_0^{R{\nu \over 2}}>$
have a pole ${1 \over |\n| }$
which is a remainder of the original divergencies.
As a consequence,
the complete perturbative result of
order $\nu^n$ requires the evaluations of terms which in ordinary
perturbation theory appear at any order $\nu^m$,
with $n\le m\le 2n$.

In spite of these peculiarities, also the
regularization method II has
proven very reliable in a series of tests. The first order
eigenenergies and eigenfunctions for the 2-anyon Hamiltonian
$H_2$ have been calculated\cite{gac} and the
results are in agreement with the
expansion in $\nu$ of the exact solutions.
Some 3 and 4 anyon eigenenergies have been
calculated\cite{cho,mit,papsix} to second
order in $\nu$, with results in agreement with the numerical solutions
obtained in Refs.\cite{verba,murthy}
and (for the states for which such exact solutions are available) with the
exact solutions found in Refs.\cite{wu,tru,cho}.
However, in the second order calculations some spurious infinities must be
neglected in order to obtain the correct results\cite{papsix}.
Even though this further regularization can be cast into a general
procedure\cite{papsix}
(the same term is neglected in all second order calculations), this
is another arbitrary manipulation required by the regularization method II.

The quantitative successes of the regularization method II
are surprising considering the apparent arbitrariness of
some of the manipulations involved.
In particular, even if the basic physical idea
(that in order to study conventional anyons the
unperturbed wave functions must have an hard-core) should be correct,
one would like to understand why the definition (\ref{eqcq}) is the right one.
For example, if one used the definition
$|\Psi_0^{R \alpha}> \equiv N_{\Psi_0,\alpha}~r^{{|\alpha|}}~ |\Psi_0>$
the $|\Psi_0^{R \alpha}>$'s would have an hard-core independently of the
value of $\alpha$,
and it is not clear for what reason
the choice $\alpha=\nu/2$
is the only one leading to the correct anyon spectra\cite{cho,gac}.

\newsubsection{$\delta$-function regularization}
For the conventional non-colliding anyons
the wave functions vanish
at the points of overlap, and
the addition of a repulsive
$\delta$-function potential to the Hamiltonian $H_N$
of a quantum mechanical $N$-anyon system has no physical consequences
(see for example Ref.\cite{manu2}),
i.e. the exact eigensolutions are unaffected by it.
One can therefore apply small-$\nu$ perturbation theory,
instead of to the original Hamiltonian $H_N$,
to the equivalent
Hamiltonian $H^{\delta}_N$, given by
\be
H_N^{\delta} \equiv
H_{N} +
2 \pi |\nu| \sum_{m < n} \delta^{(2)}({\bf r_{n} - r_{m}}) ~.
\label{hndelgen}
\ee

For our 2-anyon in harmonic potential problem
the ``regularized Hamiltonian" is
\be
H^\delta_2 \equiv
- {1 \over r} \partial_{r} (r \partial_{r})
- {1 \over r^2}  \partial_{\phi}^{2} + r^2
-{ 2i\n \over  r^2}\partial_{\phi}
 + 2 \pi |\nu| \delta^{(2)}({\bf r})
+ { \nu^2 \over r^2}
=H_2 + 2 \pi |\nu| \delta^{(2)}({\bf r})
{}~.
\label{hamdel}
\ee

\noindent
Although the (exact) $H^\delta_2$-eigenproblem is completely
equivalent to the
$H_2$-eigenproblem (they have precisely the
same eigensolutions), $H^\delta_2$ is more suitable for perturbation
theory;
in fact, the added $\delta$-function potential leads to
divergencies which exactly cancel those introduced by the $\nu^2 / r^2$
term,
rendering finite the results of bosonic end perturbation theory\cite{pap9}.

I shall illustrate the mechanism that leads to these cancellations
in Sec.IV; here I want to discuss the physical interpretation
of the $\delta$-function regularization procedure.
Within the context of perturbation theory in quantum mechanics,
the addition to the original Hamiltonian of a
term $2 \pi |\nu| \delta^{(2)}({\bf r})$
can simply be interpreted as an expedient to
implement the hard core
boundary conditions in the perturbative calculations\footnote{More
on contact interactions and boundary
conditions can be found in Refs.\cite{jakdelta,papstramitejak}.}.
One can also see, following the analysis in Refs.\cite{manu,jakdelta},
that such a $\delta$-function potential is naturally
induced by a procedure of perturbative renormalization of anyon
quantum mechanics, in complete analogy with the structure of the
regularization and renormalization procedures used in field theory.

Indeed, in the study of the non-relativistic field theories
that correspond to
our quantum mechanical problem it has been shown\cite{lozanob,bakber} that
a quartic contact interaction
(the field theoretical analog of a $\delta$-function potential) is necessary
for renormalization.
In these field theory contexts one can also see that at some critical
values of the quartic contact interaction strength the theory is finite
(and preserves classical conformal invariance).
In the framework of a perturbative renormalization of anyon
quantum mechanics a similar interpretation can be given of the
choice\footnote{Note, however, that both in field theory and in quantum
mechanics these critical values of the interaction strengths are
only determined up to sign ambiguities (see for example Ref.\cite{bakber}),
and the sign-choice can only be made by comparing perturbative results
with the exact results one wants to reproduce.}
$2 \pi |\nu|$ for the coefficient of the $\delta$-function term which is
necessary in order to reproduce the exact results.
The critical value $2 \pi |\nu|$ can be shown\cite{bakber} to be the one that
implements the boundary conditions appropriate for the conventional
non-colliding anyons
in the perturbation theory.

In summary, our $\delta$-function regularization procedure has
indeed a
rather transparent physical interpretation, and makes contact
with other interesting problems of theoretical physics.

\newsection{Relations between Regularization Procedures}
In this section I will individuate some relations between the regularization
procedures discussed in the preceding section.
Again for definiteness and simplicity my discussion is specialized to the
2-anyon in external harmonic oscillator potential problem.

Let me start by looking at the relation between the regularization
method I and the $\delta$-function regularization.
This is most simply seen by observing that\cite{sen}
\be
{\tilde{H}_2^\beta+\tilde{H}_2^{\beta +} \over 2}
\sim  - {1 \over r} \partial_{r} (r \partial_{r})
- {1 \over r^2}  \partial_{\phi}^{2} + r^2
 -{ 2i\n \over  r^2}\partial_{\phi}
+ 2 \pi |\beta| \delta^{(2)}({\bf r})
 ~,\label{hamsen}
\ee

\noindent
for $\beta \! = \! \nu$ includes the regulating
$\delta$-function potential
$2 \pi |\nu| \delta^{(2)}({\bf r})$.
This suggests that the regularization
method I might be equivalent to the $\delta$-function regularization.
Indeed, following the analysis of Ref.\cite{ouvnew} one finds that
$H_2^{\delta}$ can be obtained from $\hv_2^\nu$ by performing a non-unitary
transformation that is the inverse of the ones used in Eq.(\ref{eqcb}), i.e.
\be
H_2^{\delta} \sim r^{|\nu|} \tilde{H}_2^\nu r^{-|\nu|}
 ~.\label{cominv}
\ee

\noindent
It is therefore not surprising that the bosonic end
perturbative approaches
based on $H_2^{\delta}$ and $\tilde{H}_2^{\nu}$ lead to consistent
results, as indicated by the calculations in the literature.

There is one more noteworthy observation
to be made about Eq.(\ref{hamsen}). In fact, this equation shows that
(due to the omission of the important $\nu^2/r^2$ term)
a bosonic end perturbation theory based on the
Hamiltonian $(\tilde{H}_2^{\nu}+\tilde{H}_2^{\nu +})/2$,
first discussed
in Ref.\cite{sen},
is not equivalent to
the $\delta$-function regularized bosonic end perturbation theory.
As already
noticed in Ref.\cite{sen},
the Hamiltonian $(\tilde{H}_2^{\nu}+\tilde{H}_2^{\nu +})/2$
can only be used to obtain first order results (as shown
in Ref.\cite{gac}, the second order
results are divergent).

Concerning the relation
between the $\delta$-function regularization and the
regularization method II, I observe that
\be
<\Psi^{(0)}_{m,k}| ~
2 \pi |\nu| \delta^{(2)}({\bf r}) ~
|\Psi^{(0)}_{n,l}> ~ = ~
\biggl[ <\u^{(0)R {\nu \over 2}}_{m,k}| ~ {\nu^2 \over r^2} ~
|\Psi^{(0)R {\nu \over 2}}_{n,l}> \biggr]_{O(\nu)}
{}~.
\label{madr}
\ee

\noindent
This suggests that the
regularization of the wave
functions $|\Psi^{(0)}_{n,l}> \rightarrow |\Psi^{(0)R {\nu \over 2}}_{n,l}>$
used in the regularization method II
is in some correspondence with the regularization of the Hamiltonian
$H_2 \rightarrow H_2 + 2 \pi |\nu| \delta^{(2)}({\bf r})$ used in the
$\delta$-function regularization.
However, it is easy to see\cite{papsix}
that (unlike the second order results of the
$\delta$-function regularization) the second order results of the
regularization method II
require additional regularization. (As I mentioned, this additional
regularization can be cast into a general procedure
that has been shown to give correct second order
results in several tests\cite{cho,mit,papsix}, but presently the situation
of the higher orders in the perturbative expansion is not clear.)
In fact, one can easily see that the regularization method II is
closer in spirit\footnote{As shown by Eq.(\ref{madr}) the matrix elements
$<\u^{(0)R {\nu \over 2}}_{m,k}| ~ {\nu^2 \over r^2} ~
|\Psi^{(0)R {\nu \over 2}}_{n,l}>$
of the regularization method II correspond to the matrix elements
$<\Psi^{(0)}_{m,k}|
2 \pi |\nu| \delta^{(2)}({\bf r})
|\Psi^{(0)}_{n,l}>$
of the hermitized version of the regularization method I,
and in both these approaches one essentially misses
second order contributions of the type
$<\u^{(0)}_{m,k}| ~ {\nu^2 \over r^2} ~
|\Psi^{(0)}_{n,l}>$.}
to the ``hermitized version'' of the regularization method I,
but is formulated
in a way that naturally suggests a general procedure of regularization
of the second order results.

\newsection{Tests of the $\delta$-function Regularization}
I now perform some direct tests
of the reliability of the $\delta$-function regularization.
I calculate
some first and second order eigenenergies and some first
order eigenfunctions, and compare them with the corresponding exact
results. I consider identical anyons in an external harmonic
oscillator potential; however, it is easy to realize that the particular
form of the potential does not play a role in the regularization
procedure. Indeed, the divergence that one is regularizing results
from the ``free-anyon part'' of the Hamiltonian.

\newsubsection{2-anyon energies and wave functions}
As a first direct test of its reliability, in this subsection I use
the $\delta$-function regularization in the evaluation of
the first and second order eigenenergies and
the first order eigenfunctions of the 2-anyon Hamiltonian $H_2$ given
in Eq.(\ref{eqbe}),
whose $\delta$-function regularized Hamiltonian $H_2^\delta$
was given in Eq.(\ref{hamdel}).
My analysis is limited to the anyonic states  whose $\nu \rightarrow 0$
limit are bosonic states with angular momentum $l=0$.
For the states with $l \ne 0$ no divergence is present to begin
with\cite{cho,comtet,gac},
and the consistency of the $\delta$-function regularization
can be verified in complete analogy with the corresponding results
obtained for the other procedures. [N.B. The
$\delta$-function potential does not contribute to the matrix elements
involving unperturbed
states with $l \ne 0$, because these states vanish
for ${\bf r} = 0$.]

Concerning the first order energies,
one easily finds
\bea
E^{(1)}_{n,0,\nu}
= <\Psi^{(0)}_{n,0}|
-{ 2i\n \over  r^2}\partial_{\phi}
 + 2 \pi |\nu| \delta^{(2)}({\bf r})
|\Psi^{(0)}_{n,0}> \nonumber\\
= <\Psi^{(0)}_{n,0}| 2 \pi |\nu| \delta^{(2)}({\bf r})
|\Psi^{(0)}_{n,0}>
= 2 |\nu|
{}~,
\label{eonecoppia}
\eea

\noindent
which is clearly in agreement with Eq.(\ref{solb}).

The first order eigenfunctions are given by
\bea
|\Psi^{(1)}_{n,0,\nu}> & = &
\sum_{m,l \ne n,0}
{<\Psi^{(0)}_{m,l}| ~
-{ 2i\n \over  r^2}\partial_{\phi}
 + 2 \pi |\nu| \delta^{(2)}({\bf r}) ~ |\Psi^{(0)}_{n,0}>
\over E^{(0)}_{n,0}-E^{(0)}_{m,l} } |\Psi^{(0)}_{m,l}>
\nonumber\\
&= & \sum_{m[\ne n]} {<\Psi^{(0)}_{m,0}| ~
2 \pi |\nu| \delta^{(2)}({\bf r}) ~
|\Psi^{(0)}_{n,0}> \over E^{(0)}_{n,0}-E^{(0)}_{m,0} }
|\Psi^{(0)}_{m,0}>
\nonumber\\
& = & - {|\n| \over 2\sqrt{\pi }} \sum_{m \ne n}
{L_{m}^{0}(r^2) \over m-n} e^{-{r^2 \over 2}}
{}~.
\label{psionecoppia}
\eea

\noindent
Using properties of
the Laguerre polynomials one can verify (with some algebra) that the
result (\ref{psionecoppia}) is in agreement with Eq.(\ref{sola}). For
example one obtains
\be
|\Psi^{(1)}_{2,0,\nu}> =
 {|\nu| \over \sqrt{\pi }}  e^{-{r^2 \over 2}}
\biggl[ {3 \over 2} - r^2 + {1 \over 4} (2\gamma - 3 + 4 ln(r))
L_{2}^{0}(r^2) \biggr]
{}~,
\label{psioneto}
\ee

\noindent
which agrees with the first order term in the expansion
in $\nu$ of $|\Psi^{exact}_{2,0,\nu}>$.

{}From (\ref{hamdel}) one sees that the second order energies
are given by
\bea
E^{(2)}_{n,0,\nu}
&=& E^{(2,a)}_{n,0,\nu} ~ + ~ E^{(2,b)}_{n,0,\nu} ~, \label{etwocoppia}\\
E^{(2,a)}_{n,0,\nu} &\equiv&
<\Psi^{(0)}_{n,0}| {\nu^2 \over r^2}
|\Psi^{(0)}_{n,0}> ~, \label{etwocoppiaa}\\
E^{(2,b)}_{n,0,\nu} &\equiv&
<\Psi^{(0)}_{n,0}| -{ 2i\n \over  r^2}\partial_{\phi}
 + 2 \pi |\nu| \delta^{(2)}({\bf r})
|\Psi^{(1)}_{n,0,\nu}> ~. \label{etwocoppiab}
\eea

\noindent
It is easy to realize that
both $E^{(2,a)}_{n,0,\nu}$ and $E^{(2,b)}_{n,0,\nu}$ are divergent, but
one can show that the final result $E^{(2)}_{n,0,\nu}$ is finite.
I illustrate the details of the mechanism of
cancellation of the infinities by following a definite
calculation: the one for $E^{(2)}_{2,0,\nu}$.
$E^{(2,a)}_{2,0,\nu}$ and $E^{(2,b)}_{2,0,\nu}$ are given by
\bea
E^{(2,a)}_{2,0,\nu} \!\!\! &=&
\!\!\! <\Psi^{(0)}_{2,0}| ~ {\nu^2 \over r^2} ~
|\Psi^{(0)}_{2,0}>  =
\nu^2 \int^{\infty}_{0} \int^{2 \pi}_{0} dr~ d\phi ~
{\exp( -r^2 ) \over \pi r}~ [L_{2}^{0}(r^2)]^2
\nonumber\\
&=& \!\!\!  \lim_{\epsilon \rightarrow 0}
\int^{\infty}_{\epsilon}  dr \, \nu^2 \,
{\exp( -r^2 ) \over r} \,
[L_{2}^{0}(r^2)]^2 \, = \,
\nu^2 \, \lim_{\epsilon \rightarrow 0}
\biggl[-2 \ln(\epsilon) -\gamma -{3 \over 2}\biggr]
\, ,
\label{etworega}
\eea

\noindent
and
\bea
E^{(2,b)}_{2,0,\nu} \!\!\! &=& \!\!\! <\Psi^{(0)}_{2,0}| ~
-{ 2i\nu \over r^2}\partial_{\phi}
 + 2 \pi |\nu| \delta^{(2)}({\bf r})
{}~ |\Psi^{(1)}_{2,0}> ~
= ~ <\Psi^{(0)}_{2,0}| ~ 2 \pi |\nu| \delta^{(2)}({\bf r}) ~
|\Psi^{(1)}_{2,0}>
\nonumber\\
\!\!\! &=& \!\!\! 2 \nu^2
\int^{\infty}_{-\infty} \int^{\infty}_{-\infty}  dr_x \, dr_y \,
\delta^{(2)}({\bf r}) \, e^{-r^2} \, L_{2}^{0}(r^2) \,
\biggl[ {3 \over 2} - r^2 + {1 \over 4} (2\gamma - 3 + 4 ln(r))
L_{2}^{0}(r^2) \biggr]
\nonumber\\
\!\!\! &=& \!\!\! \nu^2 \, \lim_{\epsilon \rightarrow 0}
\biggl[2 ln(\epsilon) + {3 \over 2} + \gamma \biggr]
{}~.
\label{etworegb}
\eea

\noindent
Note that I introduced a cut-off $\epsilon$
(which will be ultimately removed by
taking the limit $\epsilon\rightarrow 0$)
in order to see the cancellation of infinities and evaluate the left-over
finite result.
In general a similar cut-off must be introduced in all the divergent
matrix elements
of $r^{-2}$ and $\delta^{(2)}({\bf r})$
by using
\be
\int^{\infty}_{-\infty} \int^{\infty}_{-\infty}  dr_x ~ dr_y
{1 \over r^2}~f(r_x,r_y) =
\lim_{\epsilon \rightarrow 0}
\int^{\infty}_{\epsilon} \int^{2 \pi}_{0} r ~ dr ~ d\phi ~
{1 \over r^2}~f(r \cos \phi,r \sin \phi)
{}~,
\label{rega}
\ee
\be
\int^{\infty}_{-\infty} \int^{\infty}_{-\infty}  dr_x ~ dr_y
\delta^{(2)}({\bf r})~f(r_x,r_y) =
\lim_{\epsilon \rightarrow 0}
f({\epsilon \over \sqrt{2}},{\epsilon \over \sqrt{2}})
{}~.
\label{regb}
\ee

{}From Eqs.(\ref{etwocoppia}), (\ref{etworega}), and (\ref{etworegb}) one
concludes
that $E^{(2)}_{2,0,\nu}=0$, and this is in agreement with Eq.(\ref{solb}).

\newsubsection{Some $N$-anyon energies}
In this subsection I calculate perturbatively from the bosonic end to
second order in the statistical parameter $\nu$\space and for
arbitrary $N$
the eigenenergies of some $N$-anyon in harmonic potential states.

The $\delta$-function regularized Hamiltonian which
describes the relative motion
of $N$ identical anyons in an harmonic potential
is given by
\be
H_N^{\delta}=H^{(0)} + \nu H_L^{(1)} + |\nu| H_\delta^{(1)} + \nu^2 H^{(2)}
{}~,
\label{hamcalc}
\ee

\noindent
where $H_0$ is the relative motion hamiltonian for N
bosons in an harmonic potential, and
\bea
H_L^{(1)} &\equiv& {1 \over 2} \sum_{m \ne n}
{1 \over  |z_n - z_m|^2} L_{n,m} ~, \nonumber\\
H_\delta^{(1)} &\equiv&
2 \pi \sum_{m < n} \delta^{(2)}(z_m - z_n ) ~, \nonumber\\
H^{(2)} &\equiv& {1 \over 4} \sum_{m\ne n, n \ne k }
\left({1 \over (z_n-z_m)(z_n^*-z_k^*)} +~ h.c. ~\right) ~; \nonumber\\
L_{n,m} &\equiv&  (z_n - z_m)
\left({\partial \over \partial z_n} -{\partial \over \partial z_m}\right)
- (z_n^* - z_m^*)
\left({\partial \over \partial z_n^*} -
{\partial \over \partial z_m^*}\right) ~.
\label{hamb}
\eea

\noindent
I am using the conventional notation $z_n \equiv r_{n}^1 + i r_{n}^2$,
$z_n^* \equiv r_{n}^1 - i r_{n}^2$.
The center of mass motion is simply a free motion, which I ignore.

The $N$-anyon states whose energies I evaluate perturbatively
are the ones which correspond, in
the limit $\nu \rightarrow 0$,
to the following bosonic states
\be
|N,\Omega> \equiv {1 \over {\sqrt {\pi^{N-1}}}} ~
exp \left[ \sum_{n=1}^{N-1} |u_n(\{ z_i \})|^2 \right]
{}~,
\label{statea}
\ee
\be
|N,+2> \equiv {1 \over \sqrt{2 ~ (N-1) ~ \pi^{N-1}}}
\left( \sum_{n=1}^{N-1} u^2_n(\{ z_i \}) \right)
exp \left[ \sum_{n=1}^{N-1} |u_n(\{ z_i \})|^2 \right]
{}~,
\label{stateb}
\ee
\be
|N,-2> \equiv |N,+2>^*  ~,
\label{statec}
\ee

\noindent
where $u_n(\{ z_i \}) \equiv (z_{1}+z_{2}+ .... + z_n
- n ~ z_{n+1})/ \sqrt{n(n+1)}$.

\noindent
$|N,\Omega>$ is the $N$-boson ground state,
and the states $|N,\pm 2>$ are in the first excited
bosonic energy level and have angular momentum $\pm 2$.

\noindent
It is easy to verify
that $E^{(1,2)}_{N,\Omega} (\nu)=E^{(1,2)}_{N,\Omega} (-\nu)$,
$E^{(1)}_{N,+2} (\nu)=-E^{(1)}_{N,-2} (-\nu)$, and
$E^{(2)}_{N,+2} (\nu)=E^{(2)}_{N,-2} (-\nu)$; therefore, I can limit
the calculations to the case $\nu >0$
without any loss of generality.

The first order energies can be easily calculated; they are given by
\bea
E^{(1)}_{N,\Omega} &=&
<N,\Omega| \nu H_L^{(1)} + |\nu| H_\delta^{(1)}
|N,\Omega> \nonumber\\
&=& {N (N-1) \over 2} <N,\Omega| 2 \pi \nu  \delta^{(2)}( z_1 - z_2)
|N,\Omega> \nonumber\\
&=& {N (N-1) \over 2} \nu
{}~,
\label{eonena}
\eea
\bea
E^{(1)}_{N,\pm 2} &=&
<N,\pm 2| \nu H_L^{(1)} + |\nu| H_\delta^{(1)}
|N,\pm 2> \nonumber\\
&=& {N (N-1) \over 2} <N,\pm 2| \nu {L_{12} \over |z_1 - z_2|^2}
+ 2 \pi \nu  \delta^{(2)}( z_1 - z_2)
|N,\pm 2> \nonumber\\
&=& {N (N-2) \over 2} \nu \pm {N \over 2} \nu
{}~.
\label{eonenb}
\eea

Concerning the evaluation of the second order energies,
let me start by noticing that from (\ref{hamcalc}) and
(\ref{hamb})
it follows that
\bea
E_{\Psi^{(0)}}^{(2)}&=& <\Psi^{(0)}| \nu^2 H^{(2)}
|\Psi^{(0)}>
+E_{\Psi^{(0)};L}^{(2)}+E_{\Psi^{(0)};\delta}^{(2)}
{}~,\label{etwopiecesa}\\
E_{\Psi^{(0)};L}^{(2)}
&\equiv & \sum_{|m> \notin D} { <\Psi^{(0)}|
\nu H_L^{(1)} + |\nu| H_\delta^{(1)} |m>
<m| \nu H_L^{(1)} |\Psi^{(0)}>
\over {E^{(0)}-E^{(0)}_m}}
{}~,\label{etwopiecesb}\\
E_{\Psi^{(0)};\delta}^{(2)} & \equiv &
\sum_{|m> \notin D} { <\Psi^{(0)}|
\nu H_L^{(1)} + |\nu| H_\delta^{(1)} |m>
<m| |\nu| H_\delta^{(1)} |\Psi^{(0)}>
\over {E^{(0)}-E^{(0)}_m}}
{}~.
\label{etwopiecesc}
\eea

\noindent
Using the symmetries of $H^{(2)}$ and of the unperturbed wave
functions (\ref{statea}), (\ref{stateb}), and (\ref{statec}),
one easily obtains
\bea
<N,\Omega| \nu^2 H^{(2)}
|N,\Omega> &=&
{\nu^2 \over 4}~N(N-1)~
\biggl[ 2 (N-2) \ln({4 \over 3}) - \gamma \biggr] \nonumber\\
&  & - \lim_{\epsilon \rightarrow 0}
\biggl[ {\nu^2 \over 4}~N(N-1)~
\ln(\epsilon)\biggr]
{}~,\nonumber\\
<N,\pm 2| \nu^2 H^{(2)} |N,\pm 2> &=& {\nu^2 \over 8} ~N
\biggl[ 9 - 4 N + 4 (N+1) (N-2) \ln({4 \over 3})
- 2 \, \gamma \, (N-2) \biggr] \nonumber\\
&  & - \lim_{\epsilon \rightarrow 0} \biggl[ {\nu^2 \over 4} \, N \,
(N-2)\, \ln(\epsilon) \biggr]
{}~,\label{nusquare}
\eea

\noindent
where the cut-off $\epsilon$ has been introduced
using (\ref{rega}).

For the states (\ref{statea}), (\ref{stateb}), and (\ref{statec})
the evaluation of $E_{\Psi^{(0)};L}^{(2)}$
and $E_{\Psi^{(0)};\delta}^{(2)}$
(which usually is only possible numerically) is
relevantly simplified by
the following results
\bea
{ 1\over |z_1-z_2|^2} L_{12} \, |N,\pm 2>
&=& {1\over 4}
\left( [C, H^{(0)}]- \pi \delta^{(2)}({z_1-z_2 \over {\sqrt 2}})
+ 1 \right)
L_{12} \, |N,\pm 2>  ~, \nonumber\\
<m| \, [C, H^{(0)}] \, |N,\Omega (\pm 2)>
&=&
<m| \, \pi \, \delta^{(2)}({z_1-z_2 \over {\sqrt 2}}) \, |N,\Omega (\pm 2)>
{}~, \cr
H^{(0)} ~ L_{12} ~ |N,\pm 2> &=& E^{(0)}_{N,\pm} ~ L_{12} ~ |N,\pm 2>
{}~, \nonumber\\
L_{12} ~ |N,\Omega> &=& 0 \, . \label{cidn}
\eea

\noindent
where $C={1\over 2}(ln( {|z_1-z_2|^2 \over 2}) + \gamma - 1)$.

\noindent
Using the properties (\ref{cidn}) one finds
\be
E_{N,\Omega;L}^{(2)}=0
{}~,\label{rescompa}
\ee
\bea
E_{N,\Omega;\delta}^{(2)} &\!=\!&
{\nu \over 2} (N^2-N)
\biggl[
<N,\Omega|  (\nu H_L^{(1)} + |\nu| H_\delta^{(1)}
C |N,\Omega>
- E_{N,\Omega}^{(1)}
<N,\Omega| C  |N,\Omega> \biggr] \nonumber\\
&\!=\!& \lim_{\epsilon \rightarrow 0}
\biggl[ {\nu^2 \over 4}~N(N-1)~
\biggl( \ln(\epsilon) + \gamma
- 2 (N-2) \ln({4 \over 3}) \biggr) \biggr]
{}~,\label{rescompb}
\eea
\bea
E_{N,\pm 2;L}^{(2)} &\!=\!&
{\nu \over 8} (N^2-N)
\biggl[ <N,\pm 2| (\nu H_L^{(1)} + |\nu| H_\delta^{(1)})
\, C \, L_{12} |N,\pm 2> \nonumber\\
&\! \!& ~~~~~~~~~~~~~~~~~~~~  - E_{N,\pm 2}^{(1)}
<N,\pm 2| C \, L_{12} |N,\pm 2> \biggr] \nonumber\\
&\!=\!& \pm {3 \over 8} \nu^2
N (N-2) \ln({3 \over 4})
+ {\nu^2 \over 16} ~N
\biggl(   5 N - 12 - 18 (N-2) \ln({4 \over 3})   \biggr)
{}~,\label{rescompc}
\eea
\bea
E_{N,\pm 2;\delta}^{(2)} &\!=\!&
{\nu \over 2} (N^2-N)
\biggl[
<N,\pm 2|  (\nu H_L^{(1)} + |\nu| H_\delta^{(1)})
\, C \, {\sum_{n=2}^{N-1} u^2_n(\{ z_i \}) \over
\sum_{k=1}^{N-1} u^2_k(\{ z_i \})} \, |N,\pm 2> \nonumber\\
&\! \!& ~~~~~~~~~~~~~~~~~~~~ - E_{N,\pm 2}^{(1)}
<N,\pm 2| C
{}~ {\sum_{n=2}^{N-1} u^2_n(\{ z_i \}) \over
\sum_{k=1}^{N-1} u^2_k(\{ z_i \})} ~
|N,\pm 2> \biggr] \nonumber\\
&\!=\!& {\nu^2 \over 16} \, N \, \lim_{\epsilon \rightarrow 0}
\biggl[   3 N - 6
+ 4 (N-2) (\gamma + \ln(\epsilon))
+ (38 N -44 - 8 N^2) \ln({4 \over 3}) \biggr] \nonumber\\
& & \pm {3 \over 8} \nu^2 N (N-2) \ln({3 \over 4})
{}~.\label{rescompd}
\eea

\noindent
Again I introduced the cut-off $\epsilon$ using (\ref{regb}).

Combining these results with (\ref{eonena}), (\ref{eonenb}),
(\ref{etwopiecesa}), and
(\ref{nusquare}) one obtains the following
final results:
\be
E_{N,\Omega}^{(1)} + E_{N,\Omega}^{(2)} =
{N (N-1) \over 2} \nu
{}~,\label{resa}
\ee
\be
E_{N,+2}^{(1)} + E_{N,+2}^{(2)} =
{N (N-1) \over 2} \nu
{}~,\label{resb}
\ee
\be
E_{N,-2}^{(1)} + E_{N,-2}^{(2)} =
{N (N-3) \over 2} \nu
+ \ln({4 \over 3})
{3 N (N-2) \over 2}
\nu^2
{}~.\label{resc}
\ee

\noindent
$E_{N,\Omega}$ and $E_{N,+2}$ are among the exactly known eigenenergies of
the $N$-anyon in harmonic potential problem (see Ref.\cite{cho}),
and they are in perfect agreement with (\ref{resa}) and (\ref{resb}).
$E_{N,-2}$ is not known exactly, but,
for the special cases $N$=3 and $N$=4,
$E_{N,-2}^{(1)} + E_{N,-2}^{(2)}$  has been calculated numerically
in Refs.\cite{verba,murthy} and perturbatively,
using the regularization method II, in Refs.\cite{mit,papsix};
Eq.(\ref{resc}), for the
corresponding values of $N$, is in agreement with those results.

\newsubsection{Additional 3-anyon energies}
The N-anyon eigenenergies calculated to second order in $\nu$
in the preceding subsection and in Ref.\cite{pap9}
all concerned bosonic end states
with particularly symmetric wave functions (besides the bosonic
symmetry under exchange of the particle indices, they were also symmetric under
permutations of the functions $u_i$).
This special form of the bosonic end wave functions leads to simplifications
that were exploited in obtaining the results for arbitrary $N$.
In order to perform a more general test of the $\delta$-function
regularization (i.e. show that the agreement with the exact results
found in the preceding subsection is not a consequence of the special
form of the wave functions considered), I now calculate
to second order in $\nu$ the eigenenergies of the two 3-anyon states
whose bosonic end wave functions are given by
\bea
&|3,+3> \equiv {1 \over {\sqrt { 24 \pi^3}}} u_2(\{ z_i \})
\left( 3 [u_1(\{ z_i \})]^2 - [u_2(\{ z_i \})]^2 \right)
exp \left[ |u_1(\{ z_i \})|^2
+ |u_2(\{ z_i \})|^2 \right] ~,& \nonumber\\
&~~~~~~~~~~~~~~~~~~|3,-3> \equiv |3,+3>^*  ~.&
\label{sta}
\eea

\noindent
(Notice that
$|3,\pm 3>$ are not symmetric under permutations of the
functions $u_i$.)
The states $|3,\pm 3>$ are in the second excited energy level and have
angular momentum $\pm 3$.

Concerning the first order energies $E_{3,\pm 3}^{(1)}$ one easily finds that
\be
E^{(1)}_{3,\pm 3} = \left( {3 \over 4}
\pm {9 \over 4} \right) \nu ~.
\label{e3uno}
\ee

It is also easy to show that the states $|3,\pm 3>$ satisfy relations
of the type (\ref{cidn}), and this allows to evaluate the second order
energies $E^{(2)}_{3,\pm 3}$ analytically.
Following a procedure completely analogous to the one discussed in the
preceding subsection, I find that
\bea
E_{3,\pm 3}^{(2)}&=& <3,\pm 3 | \nu^2 H^{(2)}
|3,\pm 3>
+E_{3,\pm 3;L}^{(2)}+E_{3,\pm 3; \delta}^{(2)}
{}~,\nonumber\\
<3,\pm 3 | \nu^2 H^{(2)}
|3,\pm 3> &=& {\nu^2} \left( 2 + {9 \over 16}
- 6 ln({4 \over 3}) \right)
{}~,\nonumber\\
E_{3,\pm 3;L}^{(2)}
&=& {\nu^2 \over 64} \left( 486 ln({4 \over 3}) - 189 \right)
\pm {\nu^2 \over 32} \left(9 - 27 ln({4 \over 3}) \right)
{}~,\nonumber\\
E_{3,\pm 3; \delta}^{(2)} &=&
{\nu^2 \over 64} \left( 6 ln({4 \over 3}) - 11 \right)
\pm {\nu^2 \over 32} \left(9 - 27 ln({4 \over 3}) \right)
{}~.
\label{proe3due}
\eea

{}From Eqs.(\ref{e3uno}) and (\ref{proe3due}) one concludes that
\bea
E_{3,+ 3}^{(1)} + E_{3,+ 3}^{(2)} \!\!&=&\!\! 3 \nu ~, \label{e3plus}\\
E_{3,- 3}^{(1)} + E_{3,- 3}^{(2)} \!\!&=&\!\! - {3 \over 2} \nu
+ {9 \over 8} \left( 3 ln({4 \over 3}) - 1 \right) \nu^2 ~.
\label{e3min}
\eea

As expected the $\delta$-function regularization
result for $E^{(1)}_{3,+3} + E^{(2)}_{3,+3}$
is in perfect agreement with the corresponding exact result
obtained in Ref.\cite{wu}.
$E^{(1)}_{3,-3} + E^{(2)}_{3,-3}$ is not known exactly, but it was calculated
numerically in Refs.\cite{verba,murthy} and perturbatively, using the
regularization method II, in Ref.\cite{mit}; Eq.(\ref{e3min}) agrees with
those results.

\newsection{Summary and Conclusion}
In the first part of the paper I have examined the three bosonic
end perturbative approaches that have been used in the study of the
anyon spectra.
I find that the $\delta$-function regularization
has the most transparent physical interpretation,
and this is very important if we want to be confident in using bosonic end
perturbation theory in the study of the presently unknown portion of the
anyon spectra.
Moreover,
as illustrated by the calculation for arbitrary number of anyons
in Sec.IV, in some instances properties of
the $\delta$-function regularized Hamiltonian
can be exploited to achieve significant simplification.

I have also presented evidence in support of the equivalence of the
regularization method I and the $\delta$-function regularization
(which was already argued for in Ref.\cite{ouvnew}).
This equivalence is confirmed by the agreement between all the results
obtained with these two methods.
The regularization method I has a less insightful physical interpretation,
but the fact that the corresponding Hamiltonian only includes 2-body
interactions can be useful in some calculations.

Concerning the regularization method II, I found that it essentially
corresponds to a less efficient (and less motivated) way to
implement the same program of the $\delta$-function regularization.
As indicated by calculations presented in Sec.IV,
the results of the regularization method II
are reliable at least to
second order in $\nu$. In particular, using the
$\delta$-function regularization,
I confirmed the important result for $E_{3,-3}$ obtained in Ref.\cite{mit},
which, combined with results of Ref.\cite{cho}, allows an
approximate analytic description of the 3-anyon in harmonic oscillator
potential ground state energy for all values of $\nu$.
These calculations of $E_{3,-3}$ also represent
the only analytic evidence for the existence
of anyonic wave functions with energies non-linearly dependent
on $\nu$ that smoothly interpolate from the bosonic
to the fermionic limit\cite{mit2}.
Now that the result of Ref.\cite{mit} has been confirmed using a more
transparent procedure, there is additional motivation for the
use of the analysis presented in Refs.\cite{cho,mit} as a guide in the
search for additional exact solutions to the $N$-anyon ($N>2$) problem.

The calculation of $E^{(1)}_{N,-2}+E^{(2)}_{N,-2}$ discussed in detail
in Sec.IV is an important result of the
$\delta$-function regularization;
in fact, it gives the only direct evidence\cite{pap9}
of the fact that there are
many-anyon eigenenergies with non-linear dependence  on the statistical
parameter $\nu$.
I also want to emphasize again\cite{mit,papsix,pap9} that the
presence of the
interesting factor $\ln (4/3)$ in the non-vanishing second order
contributions to the energies (see for
example Eqs.(\ref{resc}) and (\ref{e3min}))
could suggest {\it Ans\"atze}
(like the one proposed in Refs.\cite{mit,papsix})
that simplify the search for new exact anyon eigensolutions.

Finally, I want to mention a possible area of investigation which might
be motivated by the results discussed in this paper, but is not aimed at
the study of anyon physics.
As indicated in Sec.III, the $\delta$-function regularization can be
cast
precisely in the form of a renormalization procedure\cite{manu,jakdelta}
like the ones
customary in the field theoretical framework.
However, whereas in (relativistic) field theory usually
interesting results can
only be obtained in perturbation theory, in anyon quantum mechanics
in several cases one has available (at least some of) the exact solutions.
In anyon quantum mechanics it is therefore sometime possible to compare
the (renormalization-requiring) perturbative results with the
exact results. These comparisons might lead to some insight in the
physics behind the renormalization and regularization procedures
necessary in field theory. Work on this subject
is now in progress\cite{gacbak}.

\vglue 0.6cm
\leftline{\twelvebf Acknowledgements}
\vglue 0.3cm
I have benefitted from conversations with D. Bak, R. Jackiw, and
C. Manuel; these I gratefully acknowledge.

\newpage
\renewcommand{\Large}{\normalsize} 
\end{document}